# PSEUDORAPIDITY SPECTRA OF RELATIVISTIC PARTICLES EMITTED IN THE Au AND Pb INDUCED REACTIONS AT HIGH ENERGIES

B.Z. Belashev[1], M.K. Suleymanov[2,3], S. Vokál[4], J. Vrláková[4], M.Ajaz[2], K.H.Khan[2], Ali Zaman[2], Z. Wazir[2].

[1]*Institute of Geology, Karelian Research Centre Russian Academy Sciences, St., Petrozavodsk, Russia*
[2]*COMSATS Institute of Information Technology, Islamabad, Pakistan*
[3]*Joint Institute for Nuclear Research, Dubna, Russia*
[4]*University of P.J. Šafárik, Košice, Slovakia*

## Abstract

The structure of the pseudorapidity spectra of charged relativistic particles with $\beta > 0.7$ measured in Au+Em and Pb+Em collisions at AGS and SPS energies are analyzed using Fourier transformation method and maximum entropy one. The dependences of these spectra on the number of fast target protons (g-particles) are studied. They show visually some plateau and "shoulder" which are at least three selected points on the distributions. The plateau seems wider in Pb+Em reactions. The existing of plateau is expected for the parton models. The maximum entropy method confirms the existence of the plateau and the shoulder of the distributions.

## 1. Introduction

Pseudorapidity distribution of the secondary particles emitted in high energy hadron-nuclear and nuclear-nuclear interactions contain significant information of the target and projectile fragmentation. The $\eta$ values are quasi invariant and defined easily ($\eta = -\ln(tg\,\theta/2)$) in the experiment via the polar angle. The last is especially important in case of the emulsion experiments. It is known that high positive $\eta$ values correspond mainly to the projectile fragments and negative ones mainly to target fragments. The central area of the $\eta$ distribution is considered as a region of particle production (pionization area, average values of the $\eta$). The plateau in the central $\eta$ - area in high energy hadron-nuclear and nuclear-nuclear experiments is very interesting for the theoretical study. The space-time evaluation of the hadronic matter produced in the central rapidity region in extreme nucleus-nucleus collisions is described for example in [1]. It was found, in agreement with previous studies, that quark-gluon plasma (QGP) is produced at a temperature $\geq$ 200-300 MeV, and that it should be survive over a time scale $\geq$ 5 fm/c. The author commented that the description relies on the existence of a flat central plateau and on the applicability of hydrodynamics.

The main goal of this paper is to search if some theoretical approaches can confirm the existence of the plateau. We discuss some methods of a posteriori increasing resolution of the spectral lines. We consider the Fourier transformation method [2] and maximum entropy one [3]. To observe more detailed information on rapidity distribution of $\pi^{\pm}$-mesons produced in $\pi^{-12}$p - and $\pi^{-12}$C -reactions at 40 GeV/c a new method was applied on bases of the Fourier algorithm (regularization of the width of the spectral lines )[2]. The complex structures were identified by this method as is shown in the Fig.1. One can see that at least two selected points are indicated by this method.

For our analysis we used the experimental data of the EMU12 experiment/CERN SPS [4] and E 863 experiment/ BNL AGS [5].

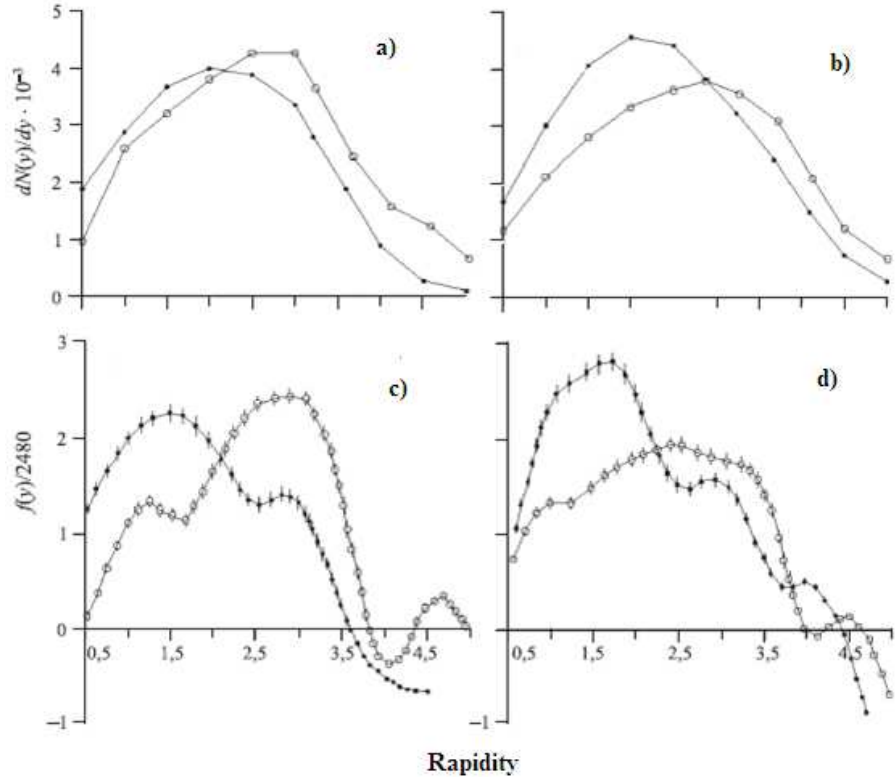

**Fig. 1** The rapidity distributions (top panels) and their Fourier transformations (bottom panels) for the $\pi^+$ - (close symbols) and $\pi^-$ - (open symbols) mesons produced in $\pi^-p$ (left panels) and $\pi^{-12}C$ (right panels) at 40 GeV/c.

## 2. Experimental data

The stacks of NIKFI BR-2 nuclear emulsions were irradiated horizontally by $^{208}$Pb beam at 158 A GeV/c (experiment EMU12 at the CERN SPS) and by $^{197}$Au beam at 11.6 A GeV/c (experiment E863 at the BNL AGS). The projectile and target nuclei are disintegrated in the interactions of relativistic nuclei. According to the geometrical picture, there are in such collisions an interaction area and nonoverlapping parts of both nuclei disintegrated to fragments due to the obtained excitation. As a result of an impact parameter variation, there appear events with the emission of secondaries of different types and energies within a wide range of multiplicity. The used emulsion method allows to measure multiplicities and angles of any charged particles and charge of projectile fragments. Secondary charged particles used in this study were classified into the following groups:

- Relativistic s particles ($N_s$), fast singly charged particles with $\beta \geq 0.7$. This group includes particles produced in the interactions (mainly pions), relativistic singly charged projectile fragments as well as those singly charged particles knocked-out from the target nucleus.

- Fast target fragments, g-particles ($N_g$), with $0.23 \leq \beta < 0.7$. They consist mainly of recoil protons from the target.

In this work we have analyzed 628 Pb+Em collisions and 1185 Au+Em collisions found by the along-the track scanning. Further details on experiments, measurements and experimental criteria can be found in [4-5].

### 3. The methods and procedures of the data analyses

Pseudorapidity distributions were considered in framework of the s = h·f + n model, where h is blurring function, n - additive noise, f – estimation of distribution which defines structure of a distribution.

Reconstruction of the structure for pseudorapidity distributions of secondary particles was done by a posteriori decreasing the widths of its components [6]. As a results of artificial decreasing of the widths at constant positions the resolution, the contrast. The methods of the Fourier-transformation [2] and the maximum entropy [3] were applied to decreasing of widths for the distribution components of the given values. Such data processing is especially evident for a case of Gaussian or Lorentz (Breit-Wigner) forms of components. In Fourier-algorithm they pass from initial distribution to its Fourier-image which is divided into the module of the Fourier-image of the components with given width and do inverse transition. The method of the maximum entropy removes given blurring function from distribution and maximizes the entropy of estimation at performance of models restrictions.

The influence of a noise to the maximum entropy method is less than for the Fourier-transformation one and could give the stable estimations at levels of additive noise 25 %. It works with narrow and wide, space-invariant and no space-invariant blurring functions [3]. The narrow blurring functions could detect more deep details of structure, the wide ones leave in estimations of distributions wide intensive components. To analyse the pseudorapidity distributions of the secondary particles and search for the existence of central plateau the maximum entropy approach seems to be more preferable.

The statistical fluctuations of the spectra were deleted using wavelet-transformation of distribution [7] at the first stage of the processing. This is equivalent to removal of high-frequency parts of signal from the spectrum. This procedure is effective at high level noise and differs considerably from the traditional averaging of histograms on counting and maintains in estimations the smoothed fluctuations. The algorithm was realized in package Wavelet Toolbox of the computer mathematics system «MATLAB» by means of a *"wdcbm"* command, using approximating *c* and detailing l factors of wavelet decomposition of distribution and parameters *alpha=1.02* and *m=l(1)* [7]. To avoid occurrence of small (on absolute values) negative values in the estimation the ones smaller than 30 were attributed to initial given histogram's data.

In the next step the components' widths were reduced assuming that the forms of components were Gaussian or Lorentz functions in the case of Fourier-algorithm and Gaussian one in the case of maximum entropy method. The Fourier-algorithm was done using the computer mathematics system «MATLAB» with the command *"FFT"* (Fast Fourier Transformation). For the Gaussian forms the following multiplicative functions of the Fourier-image were used *"$\exp(0.0001 \cdot (w_i/\Delta w)^2)$; $\exp(0.0002 \cdot (w_i/\Delta w)^2)$"*, and for the Lorentz form the function *"$\exp[0.04 \cdot |w_i/\Delta w| - 0.0001 (w_i/\Delta w)^2]$"*, where $w_i$ is cyclic frequency of i-th counting of the Fourier-image of distribution, $\Delta w$ - a step of the cyclic frequency.

In the maximum entropy method the computing program «MEMFR» [8] and the blurring function $h(i,j) = \exp[-(0.078 \cdot (\eta(i) - \eta(j))/\Delta\eta)^2]$ were taken in next analysis of the pseudorapidity spectra. For the convenience of interpretation of the processing results the maximum from the estimations were fetched to the maximal value of the initial histogram.

### 4. Experimental results

Usually the rapidity (or pseudorapidity) spectra of relativistic particles emitted at high energies are smooth without any singularities. Apparently it is not simple to extract any information on jet's production and on others effects. The pseudorapidity distributions of charged particles with β > 0.7 measured in Au+Em and Pb+Em collisions at AGS and SPS energies were analyzed and are shown on the Fig.2.

The vertical lines indicate the points which could be correspond to the bounders of plateaus and some "shoulders". The values of the η correspond to the plateaus and shoulders were taken visually. So we can say that visually the distributions contain at least three selected points.

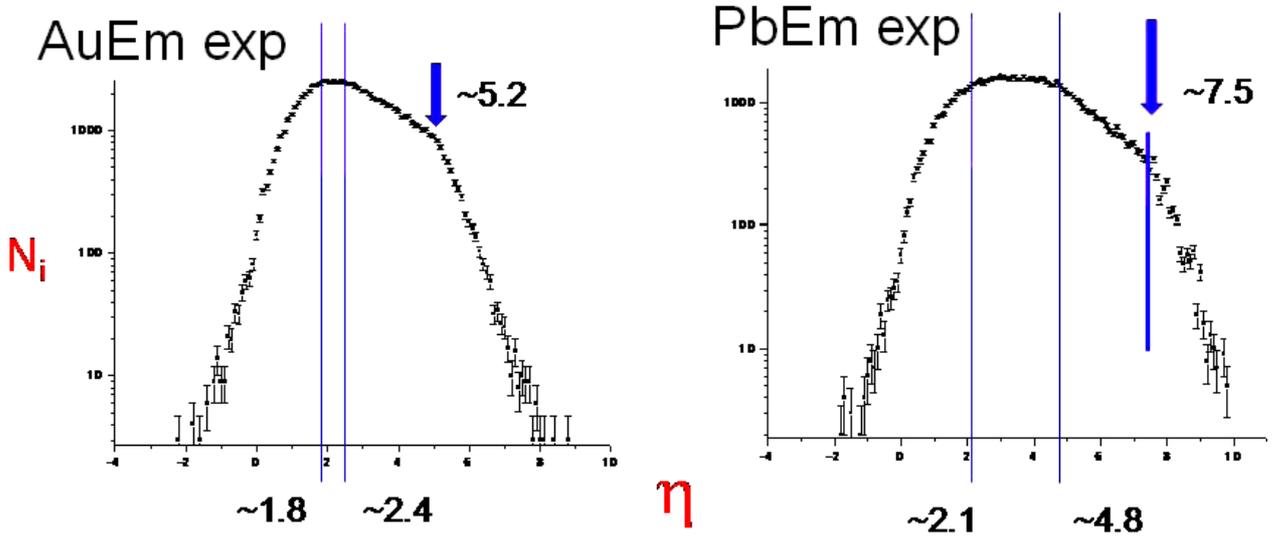

**Fig.2:** The pseudorapidity distributions for the secondary charged particles emitted in: the Pb+Em reactions at SPS energies [4] (right) and in the Au+Em reactions at AGS energies [5] (left).

The spectra were analyzed using different Fourier transformations (see section 3). This method can indicate some fluctuations but it is impossible to fix any selected points. For illustration in Fig.3 one can see the picture after applying the Fourier algorithm with multiplicative function "$exp[0.04 \cdot |w(i)/\Delta w| - 0.0001(w(i)/\Delta w)^2]$".

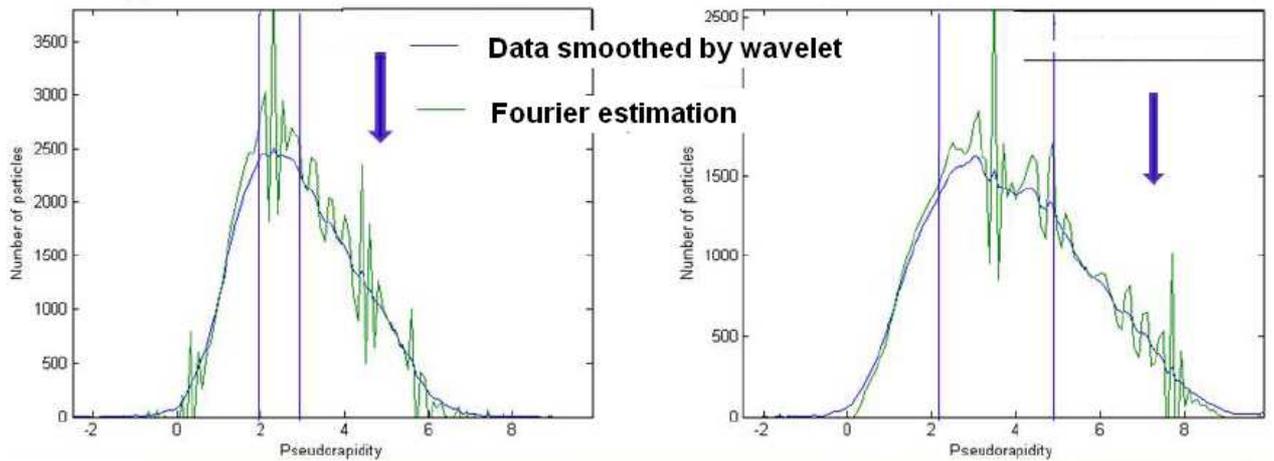

**Fig.3:** The pseudorapidity distributions after applying one of the Fourier transformation (see text) for the secondary charged particles emitted in the Pb+Em reactions at SPS energies [4] (right) and in the Au+Em reactions at AGS energies [5] (left).

Next figure (Fig. 4) shows the results after applying the maximum entropy method. One can see that the method indicates several selected points. Some of them correspond with visually observed ones.

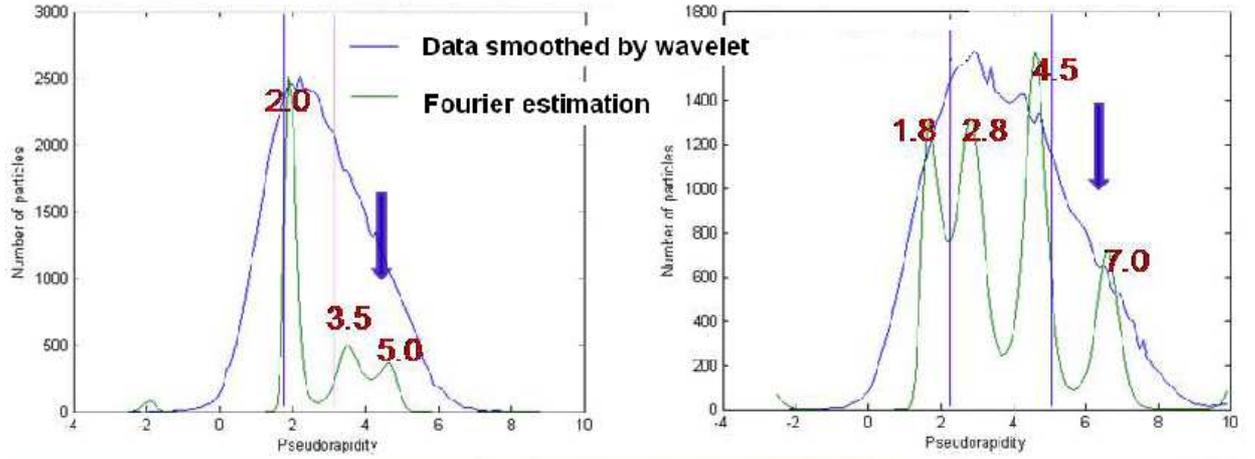

**Fig.4** The pseudorapidity distributions after applying the maximum entropy method (see text) for the secondary charged particles emitted in the Pb+Em reactions at SPS energies [4] (right) and in the Au+Em reactions at AGS energies [5] (left).

All above mentioned methods were applied to the pseudorapidity spectra of charged relativistic particles with $\beta > 0.7$ measured in Au+Em and Pb+Em collisions at AGS and SPS energies with different number of g-particles. Non statistical significant points were selected by the Fourier transformation method. On the other side the maximum entropy approach can determine some selected points of these distributions. The result of selected pseudorapidity values are shown in Table 1 together with the visual observed ones. One can see that the numbers of selected points are different for Au+Em and Pb+Em reactions, for Pb+Em it is greater than for the Au+Em. The results of this method are closed to visual observed values. The points connected to "shoulder" disappeared for the Au+Em reactions in cases of Ng=15-19 and Ng≥ 20. This would be connected with suppression of the stripping effect in the central collisions. The presented data confirm the existing plateau in the central region. The selected values of pseudorapidity would be studied using model data.

Tab.1: The pseudorapidity values for Au+Em and Pb+Em interactions determined visually (2nd line of table, see Fig.2) and for interactions with different number of Ng-particles obtained by the maximum entropy method.

| Ng \ Reactions | Au+Em | | | Pb+Em | | | |
|---|---|---|---|---|---|---|---|
| All Ng experiment | 1.8 | 2.4 | 5.2 | 2.1 | - | 4.8 | 7.5 |
| All Ng method | 2.0 | 3.5 | 5.0 | 1.8 | 2.8 | 4.5 | 7.5 |
| Ng=0-1 method | 2.0 | - | 5.0 | 1.8 | - | 4.5 | 7.0 |
| Ng=2-4 method | 1.8 | 3.2 | 5.2 | 2.2 | - | 4.5 | 7.0 |
| Ng=5-9 method | 2.0 | - | 4.0 | 2.0 | - | 4.2 | 7.0 |
| Ng=10-14 method | 2.0 | - | 4.5 | 2.0 | - | 4.2 | 7.0 |
| Ng=15-19 method | 1.8 | - | - | 1.8 | 3.2 | 5.0 | 7.0 |
| Ng≥ 20 method | 1.8 | 3.7 | - | 2.2 | - | 4.5 | 7.0 |

## Conclusions

1. The methods of a posteriori increase the resolution of the spectral lines – Fourier transformation and maximum entropy methods were applied to obtain more detail information on a structure of pseudorapidity spectra of relativistic particles emitted in the Au and Pb induced reactions at high energies.
2. The central plateau and shoulder are seen visually on the pseudorapidity spectra.
3. Both methods demonstrate complex structure of the spectra.
4. Non significant pseudorapidity values of the particle distributions have been obtained by Fourier transformation method.
5. The maximum entropy method can confirm the existence of some $\eta$ values which would be connected with the boundary values of the central plateau and the values of the $\eta$ shoulder.
6. This method has extracted some selected values of pseudorapidity which could not be observed visually; the numbers of selected points are different for Au+Em and Pb+Em reactions, for Pb+Em the number is greater than for the Au+Em.
7. The results of analysis confirm the existence plateau in the central region.

## Acknowledgments

Financial supports from the Scientific Agency of the Ministry of Education of the Slovak Republic and the Slovak Academy of Sciences (Grant No. 1/0080/08) and from the HEC Pakistan are cordially acknowledged.